\documentclass[floats,twocolumn,prd,preprintnumbers,shortbibliography,10pt]{revtex4-1}
\usepackage{amsfonts,amsmath,graphicx,epsfig,amssymb}
\usepackage[colorlinks=true,urlcolor=blue,linkcolor=blue,citecolor=blue]{hyperref}
\usepackage{slashed,color,amsmath,amssymb,mathrsfs}
\usepackage{amsfonts,epsfig,amssymb}
\usepackage{graphicx}
\usepackage{hyperref}
\usepackage{array}
\usepackage{accents}
\usepackage{tensor}
\usepackage{footmisc}
\usepackage{longtable}
\usepackage{booktabs}
\usepackage{multirow}
\usepackage{bm}
\usepackage{subfigure}
\usepackage{gensymb}

\newcommand{\Jp}{J/\psi}
\newcommand{\dd}{\mathrm{d}}

\allowdisplaybreaks

\begin{document}
	\title[]{On the quantum numbers of the $X(1880)$ }
	\author{Qin-He Yang$^{1,2,3}$}
%    \email{yqh@hnu.edu.cn}    
	\author{Ling-Yun Dai$^{1,4}$}
        \email{dailingyun@hnu.edu.cn}
%        \author{Johann Haidenbauer$^{4}$}
%       \email{j.haidenbauer@fz-juelich.de}
        \author{and Ulf-G. Mei{\ss}ner$^{2,3,5}$}
        \email{meissner@hiskp.uni-bonn.de}
	\affiliation{$^{1}$ School for Theoretical Physics, School of Physics and Electronics, Hunan University, Changsha 410082, China}
	\affiliation{$^{2}$ Helmholtz Institut f\"ur Strahlen- und Kernphysik and Bethe Center for Theoretical Physics, Universit\"at Bonn, D-53115 Bonn, Germany}
\affiliation{$^{3}$Institute for Advanced Simulation (IAS-4), Forschungszentrum J\"ulich, D-52425 J\"ulich, Germany}
\affiliation{$^{4}$ Hunan Provincial Key Laboratory of High-Energy Scale Physics and Applications, Hunan University, Changsha 410082, China}
\affiliation{$^{5}$Peng Huanwu Collaborative Center for Research and Education, International Institute for Interdisciplinary and Frontiers,
Beihang University, Beijing 100191, China}
\date{\today}
\begin{abstract}
We study the properties of the $X(1880)$, the structure around the $\bar{p}p$ threshold that appears in the $3(\pi^+\pi^-)$ invariant mass spectrum in the decay process of $J/\psi \to \gamma 3(\pi^+\pi^-)$.  Nucleon-antinucleon rescattering is taken into account in our analysis, and the decay amplitude of $J/\psi \to \gamma 3(\pi^+\pi^-)$ can be obtained by the distorted wave Born approximation. With these amplitudes, we analyze the contributions to the $X(1880)$ from different partial waves. Our analysis suggests that the $X(1880)$ should be isoscalar $0^{-+}$, and it is generated by the threshold behavior. 
%%%
\end{abstract}

\maketitle

\section{Introduction}
\label{Sec:I}
Over half a century, the quark model is believed to be the fundamental template for describing hadrons, which are composed of three quarks or a pair of quark-antiquark~\cite{Gell-Mann:1964ewy,Zweig:1964ruk}. With the advent of Quantum Chromodynamics, it became clear that other types of bound states should exist. In particular, the carriers of the strong force, 
the gluons, can form the hadrons in terms of hybrid and glueball, which is a kind of matter that can truely be called exotic. 
For recent studies addressing the role of gluons as constituents of hadrons, see, for example, Refs.~\cite{Bass:2018xmz,Ketzer:2019wmd,Vadacchino:2023vnc,Shepherd:2016dni,JPAC:2018zyd}. 
Recently, a candidate for glueball was reported by the BESIII collaboration in Ref.~\cite{BESIII:2023vvr}, named X(1880). 
It is found in the $3(\pi^+\pi^-)$ invariant mass spectrum in the decay process of $J/\psi \to \gamma 3(\pi^+\pi^-)$, with much higher statistics than the earlier measurement~\cite{BESIII:2013sbm}. Its mass and width are $M=1882.1\pm1.7\pm0.7$~MeV and $\Gamma=30.7\pm5.5\pm2.4$~MeV. The $X(1880)$ attracts huge attention from the community, as it reveals a new structure of hadron if the glueball nature is confirmed, see for example Refs.~\cite{Salnikov:2023ipo,Karliner:2024cql,Xiao:2024jmu,Ma:2024gsw,Niu:2024cfn,Ortega:2024zjx,Jia:2024ybo}. In lattice QCD some research focuses on the nature of glueballs and predicts their mass, e.g., the pseudoscalar glueball with a mass around 2.6 GeV~\cite{Sun2018,Sakai2023}. 
%%%%
Also, the $X(1880)$ is found near the proton-antiproton threshold, which is of great interest as there is excellent progress in the relevant phenomenology, e.g,  threshold enhancement \cite{Belle:2002bro,Belle:2002fay,BES:2003aic,BaBar:2013ves,BESIII:2017hyw,BESIII:2019nep,BESIII:2020ktn,BESIII:2020uqk,SND:2022wdb}, resonance structures \cite{BES:2005ega,Belle:2013jng,Belle:2012uhr,BESIII:2015xco,BESIII:2018dim,BESIII:2023vvr}, and oscillations of the time-like electromagnetic form factors \cite{BESIII:2022rrg,Milstein:2022tfx,Lin:2021xrc,BESIII:2023ioy}. Following these experimental measurements, there are many theoretical works \cite{Cao:2021asd,Qian:2022whn,Kang:2015yka,Haidenbauer:2014kja}.

Quantum numbers are basic for studying the properties of a particle. 
%In Ref.~\cite{BESIII:2023vvr}, another resonance, the $X(1840)$ or the so called $X(1835)$, is also reported with the resonance information $M=1832.5\pm3.1\pm2.5$~MeV and $\Gamma =80.7\pm5.2\pm7.7$~MeV. 
%%%%%%%%%%%%%%%%%%%%%%%%%
However, since there are many final states in the decay process of $J/\psi \to \gamma 3(\pi^+\pi^-)$, it is an arduous task for experimentalists performing a partial wave analysis to fix the quantum numbers of the X(1880). In this letter, we propose a simplified partial wave analysis to select the quantum numbers. 
%%%%
Since the $X(1880)$ is just around the $\bar{p}p$ threshold, a two-step process should work well for the interaction dynamics, namely $J/\psi \to \gamma \bar{N}N\to \gamma 3(\pi^+\pi^-)$. 
The $\bar{N}N$ re-scattering plays an essential role in final-state interaction, such as in the study of the electromagnetic form factors of proton and neutron in time-like region \cite{Yang:2022qoy,Yang:2024iuc}. A series of comprehensive partial wave amplitudes of $\bar{N}N$ scattering has already been obtained using chiral effective field theory (ChEFT) up to next-to-next-to-next-to-leading order (N$^3$LO) \cite{Dai:2017ont}. 
With these amplitudes, one can construct the $J/\psi \to \gamma 3(\pi^+\pi^-)$ decay partial wave amplitudes through the distorted wave Born approximation (DWBA)~\cite{Kang:2015yka,Dai:2018tlc,Dedonder:2018ulo}. 
%%%
One by one, the partial wave amplitudes obtained can be applied to describe all the relevant experimental data sets, the invariant mass spectra of $J/\psi\to \gamma\bar{p}p$ and $J/\psi\to\gamma3(\pi^+\pi^-)$ and the cross-section of $p\bar{p}\to 3(\pi^+\pi^-)$. 
The quantum numbers of the $X(1880)$ can be classified from the quality of fitting each partial wave amplitude to the data. 
Notice that the interference between different partial waves will not have such a distinct character as that of pure waves. The reason is that around the threshold, partial waves have a threshold factor, $p^L$. As an example, the square of the partial wave amplitudes will be either strongly suppressed or not for $P$- and $S$- waves, respectively.

\section{Formalism}
\label{Sec:II}
As discussed above, the amplitude of the decay process $J/\psi \to \gamma 3(\pi^+\pi^-)$ is obtained through $J/\psi\to\gamma N\bar{N}\to\gamma 3(\pi^+\pi^-)$ \cite{Yang:2022kpm}. 
The amplitude can be fixed well by a combined analysis on the three processes, $J/\psi \to \gamma p\bar{p}$, $\bar{p}p\to 3(\pi^+\pi^-)$, and $J/\psi \to \gamma 3(\pi^+\pi^-)$. They are solved by a set of equations established via the DWBA method,
\begin{eqnarray}
F_{1}(Q)\!&=&\! A^0_{1}(p')\!+\!\int_0^\infty\!\frac{\dd k k^2}{(2\pi)^3} \!A^0_{1}(k)\!\frac{1}{Q\!-\!2E_k\!+\!i\epsilon}\!T(k,\!p';\!E_{p'}\!) ,\nonumber\\
F_{2}(Q)\!&=&\! A^0_{2}(p)\!+\!\int_0^\infty\!\frac{\dd k k^2}{(2\pi)^3} \!T(p,\!k;\!E_k\!)\!\frac{1}{2E_k\!-\!Q\!+\!i\epsilon}\!A^0_{2}(k)\,.\nonumber\\
F_{3}(Q)\!&=&\! A^0_{3}(Q)\!+\!\int_0^\infty\!\frac{\dd k k^2}{(2\pi)^3} F_{1}(E_k)\!\frac{1}{Q\!-\!2E_k\!+\!i\epsilon}\!A^0_{2}(k)\,,\nonumber\\ \label{Eq:AmpF}
\end{eqnarray}
where the subscripts \lq$1,2,3$' represent the processes of $\Jp\to\gamma N\bar{N}$, $N\bar{N}\to \gamma3(\pi^+\pi^-)$, and $\Jp\to\gamma3(\pi^+\pi^-)$, respectively. $T$ is the partial wave scattering amplitude of $N\bar{N}\to N\bar{N}$. $Q$ is the $3(\pi^+\pi^-)$ or $N\bar{N}$ invariant mass for the different processes, $p$, $p'$ are the center-of-mass momenta of initial and final $N\bar{N}$ system, and $E$ is the energy of the nucleon, respectively. 
Following earlier works \cite{Kang:2013uia,Dai:2017ont}, the transition amplitudes and annihilation potential in Eq.~\eqref{Eq:AmpF} are parameterized as
\begin{eqnarray}
 A^{0,S}_{1,2}(p)&=&\tilde{C}^{S}_{1,2}+C^{S}_{1,2}p^2+D^{S}_{1,2}p^4\,,\nonumber\\
 A^{0,P}_{1,2}(p)&=&C^{P}_{1,2}p+D^{P}_{1,2}p^3\,,\nonumber\\
 A^{0,D}_{1,2}(p)&=&D^{D}_{1,2}p^2\,,\nonumber\\ 
 A^0_{3}(Q)&=&\tilde{C}_{3}+C_{3}Q\,. \label{Eq:Bron}
\end{eqnarray}
The superscripts $S,P,D$ are short for different partial waves. 
The parameters in the annihilation potential are taken as real numbers to keep the same formalism as that of Ref.~\cite{Kang:2013uia,Dai:2017ont}.
For $A^0_{3}$ ($\Jp\to\gamma3(\pi^+\pi^-)$), it is written in a general formalism with lowest orders of momentum.
The $\bar{N}N$ scattering amplitude $T$ is solved by the Lippmann-Schiwinger equation (LSE) \cite{Dai:2017ont}, 
\begin{eqnarray}
& &T_{N\bar{N}\to N\bar{N}}(p',p;E_p)=V_{N\bar{N}\to N\bar{N}}(p',p)+\int_0^\infty \frac{\dd k k^2}{(2\pi)^3}\nonumber\\
& &\times V_{N\bar{N}\to N\bar{N}}(p',k)\frac{1}{2E_p-2E_k+i\epsilon}T_{N\bar{N}\to N\bar{N}}(k,p;E_p) \,, \nonumber\\ \label{Eq:LS}
\end{eqnarray}
with the potentials given by ChEFT up to N$^3$LO~\cite{Dai:2017ont}. 
%%%%%
Here, we will restrict our analysis to the region below $T_{\rm lab}\lesssim 120$~MeV, corresponding to $p_{\rm lab}\lesssim 490$~MeV. 
Notice that in this low-energy region, the $N\bar{N}\to N\bar{N}$ solution fits well the phase shifts of all the partial waves obtained from the partial wave analysis (PWA) \cite{Zhou:2012ui}. Further, the PWA  is based on a large number of results on differential cross sections which allows to pin down the partial waves~\footnote{In Ref. \cite{Timmermans:1995xb}, the authors took $pp$ scattering as an example and fitted the data both with and without spin observables. They found that the resulting partial-wave amplitudes from these two approaches were in close agreement. For the tensor combination of the $^3P$ waves at 100 and 210 MeV, the differences were only 0.5\% and 2\%, respectively. Also, they found that the situation in $N\bar{N}$ PWA is comparable to the early $NN$ PWA results.}. In particular, this makes it possible to separate the spin-orbit and tensor effects. Nevertheless, it should be pointed out that the specific solution we used is one of the best choices \cite{El-Bennich:2008ytt,Zhou:2012ui,Dai:2017ont,Xiao:2024jmu} based on the current experimental data.  The $N\bar{N}$ scattering lengths of the lowest partial waves are listed in Table~3 of Ref.~\cite{Dai:2017ont}. For example, one has $a_{^{11}S_0}=-0.20-1.23i$~fm and $a_{^{31}S_0}=1.05-0.58i$~fm \cite{Dai:2017ont}, using the notation $a_{^{2I+1, 2S+1}L_J}$. These results are close to others, e.g., Ref.~\cite{Kang:2013uia}. The effective range parameters are given by Ref.~\cite{Dai:2017ont}, where one has $r_{^{11}S_0}=-3.98-11.03i$~fm and $r_{^{31}S_0}=0.65-2.08i$~fm, in terms of the notation $r_{^{2I+1, 2S+1}L_J}$. Especially, the lowest partial waves, $^1S_0$, $^3P_0$, $^3P_1$, $^1D_2$, and $^3P_2$, do not show a resonance-like behavior in the region of their most substantial variation. 
This makes all the low-energy constants (LECs) from ChEFT unchanged when the $T_{N\bar{N}\to N\bar{N}}$ is taken into Eq.~\eqref{Eq:AmpF} to produce the $J/\psi\to \gamma p \bar{p}$ and $N\bar{N}\to3(\pi^+\pi^-)$ amplitudes, leaving only a few parameters from the Born terms to be fixed, i.e., $\tilde{C}_i$, $C_i$, and $D_{i}$, see Eq.~\eqref{Eq:Bron}.

With the amplitudes obtained in Eq.~\eqref{Eq:AmpF}, one can analyze the $3(\pi^+\pi^-)$ and $p\bar{p}$ invariant mass spectra for the decay processes of $J/\psi\to\gamma3(\pi^+\pi^-)$ and $J/\psi\to\gamma p\bar{p}$, and the cross section of the scattering process of $p\bar{p}\to 3(\pi^+\pi^-)$. 
%%%%
According to parity and charge conjugation conservation, the permitted partial waves are $0^{-+}$, $0^{++}$, $1^{++}$, $2^{-+}$, and $2^{++}$ in the $3(\pi^+\pi^-)$ system for the $J/\psi$ radiative decay, ignoring the higher partial waves with total angular momentum $J>2$. Correspondingly, one needs to consider the $^1S_0$, $^3P_0$, $^3P_1$, $^1D_2$ and $^3P_2$ waves in the $N\bar{N}$ system. 
One can take each partial wave individually to fit all the datasets of  invariant mass spectra and cross sections. The partial wave that best describes all these data will be recognized as contributing to the structure around the $\bar{p}p$ threshold most.

%%%%%%
On the experimental side, there are sufficient data points (74) in the energy region of $p_{\rm Lab}\lesssim 490$ MeV, supporting the possibility of selecting the quantum numbers. 
There are 3 data points on the cross section of $p\bar{p}\to 3(\pi^+\pi^-)$~\cite{Sai:1982dv,OBELIX:1996pze,Klempt:2005pp}, 44 data points on the invariant mass spectra of $3(\pi^+\pi^-)$ in $J/\psi\to\gamma3(\pi^+\pi^-)$ (31 data points with high statistics) ~\cite{BESIII:2013sbm,BESIII:2023vvr}, and 27 data points on the $\bar{p}p$ invariant mass spectra in $J/\psi\to\gamma p\bar{p}$ (12 data points with high statistics) ~\cite{BES:2003aic,BESIII:2011aa}. 
%%%%
In contrast, we have 9, 7, and 5 parameters  for the $S$-, $P$-, and $D$-wave cases, including the ones shown in Eq.~(\ref{Eq:Bron}) and the cutoff. Also, there are five normalization factors for fitting the invariant mass spectra, in lack of the detection efficiency from the experimental measurements. Note that the cutoff is not treated as a free parameter, though we examine the impact of its different values following Refs.~\cite{Dai:2017ont,Yang:2022kpm}.
Notice that the first parameter of the Born term in $J/\psi\to\gamma \bar{p}p$ has been fixed to be one, $\tilde{C}_1^{S}=C_1^{P}=D_1^{D}=1$ as it is multiplied with the normalization factor for the detection efficiency.

\section{Results and discussions}\label{Sec:III}
\begin{figure}[htpb!]
 \centering
 \includegraphics[width=0.99\linewidth]{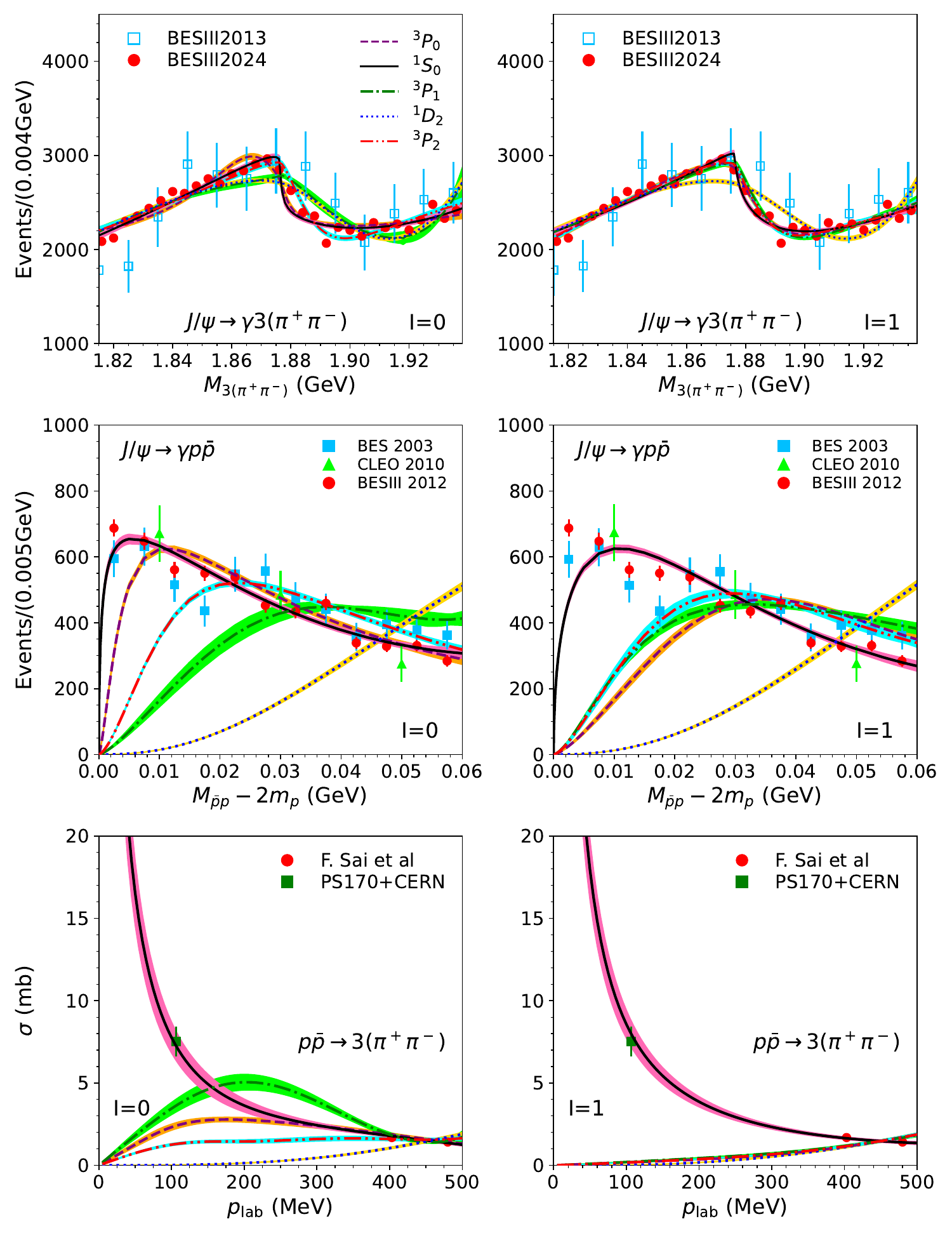}
 \caption{Comparison of N$^3$LO fit results for different partial waves. The fit results corresponding to  the $^1S_0$, $^3P_0$, $^3P_1$, $^1D_2$ and $^3P_2$ partial waves are indicated by the black solid, purple dashed, green dot-dash, blue dot, and red dot-dash-dash lines, respectively.  The data are taken from Refs.~\cite{BESIII:2023vvr,BESIII:2013sbm,Sai:1982dv,OBELIX:1996pze,Klempt:2005pp,CLEO:2010fre,BES:2003aic,BESIII:2011aa}. The cut-off is chosen as $R=1.0$~fm. In the $J/\psi\to \gamma 3(\pi^+\pi^-)$ and $J/\psi\to \gamma p\bar{p}$ decay processes, the datasets other than BESIII 2024 \cite{BESIII:2023vvr} and BESIII 2012 \cite{BESIII:2011aa} are rescaled so that all of them can be plotted together.  } 
 \label{fig:3pipi4}
\end{figure}

As discussed above, we consider the lowest partial waves, $^3P_0$ ($0^{++}$), $^1S_0$ ($0^{-+}$), $^3P_1$ ($1^{++}$), $^1D_2$ ($2^{-+}$) and $^3P_2$ ($2^{++}$), with both isoscalar and isovector channels. The fit quality for each partial wave is shown in Fig.~\ref{fig:3pipi4}, and the parameters for the different partial waves and isospins are listed Table~\ref{tab:para}.
%%%==============================================
\begin{table}
{\footnotesize	
	\centering
	\begin{tabular}{@{} c |  c  | @{} c  c  c c  c @{}}
		\hline\hline
		               PW                & $I$                & $^1S_0$     & $^3P_0$     & $^3P_1$     & $^1D_2$     & $^3P_2$     \\ \hline
		    \multirow{2}{*}{$C_{1}$}     & $0$                  & $-$7.67(15)    & --          & --          & --          & --          \\
		                                 & $1$                  & $-$18.92(53)   & --          & --          & --          & --          \\
		    \multirow{2}{*}{$D_{1}$}     & $0$                  & 44.37(95)      & $-$4.76(13)    & $-$14.28(07)   & --          & $-$5.86(21)    \\
		                                 & $1$                  & 50.00(1.26)      & $-$7.30(05)    & $-$10.98(25)   & --          & $-$9.36(07)    \\ \hline
		\multirow{2}{*}{$\tilde{C}_{2}$} & $0$                  & 0.02(01)       & --          & --          & --          & --          \\
		                                 & $1$                  & 0.01(01)       & --          & --          & --          & --          \\
		    \multirow{2}{*}{$C_{2}$}     & $0$                  & 0.25(03)       & $-$0.12(01)    & 0.30(01)       & --          & $-$0.25(02)    \\
		                                 & $1$                  & $-$0.01(04)    & $-$1.39(03)    & $-$1.27(04)    & --          & $-$1.49(04)    \\
		    \multirow{2}{*}{$D_{2}$}     & $0$                  & $-$2.53(08)    & 0.62(05)       & $-$6.52(04)    & 0.61(02)       & 3.27(08)       \\
		                                 & $1$                  & 1.32(30)       & 15.86(13)      & 50.00(3.29)      & 0.60(02)       & 27.99(22)      \\ \hline
		\multirow{4}{*}{$\tilde{C}_{3}$} & \multirow{2}{*}{$0$} & 0.75(01)       & $-$0.42(01)    & $-$1.24(01)    & $-$0.20(01)    & 0.22(03)       \\
		                                 &                        & $-$1.20(05)$i$ & +0.76(01)$i$   & $-$1.71(01)$i$ & $-$0.18(01)$i$ & +0.21(03)$i$   \\
		                                 & \multirow{2}{*}{$1$} & 0.13(04)       & 2.15(01)       & 7.78(21)       & $-$0.20(01)      & 4.95(02)       \\
		                                 &                        & +0.20(03)$i$   & +1.61(05)$i$   & +6.90(23)$i$   & $-$0.16(01)$i$   & +3.22(11)$i$   \\
		    \multirow{4}{*}{$C_{3}$}     & \multirow{2}{*}{$0$} & $-$0.57(01)    & 0.27(01)       & 0.93(01)       & 0.14(01)       & $-$0.20(02)    \\
		                                 &                        & +0.66$i$(03)   & $-$0.36(01)$i$ & +0.99(01)$i$   & +0.11(01)$i$   & $-$0.18(02)$i$ \\
		                                 & \multirow{2}{*}{$1$} & $-$0.09(03)    & $-$1.45(01)    & $-$6.14(14)    & 0.14(01)       & $-$3.45(02)    \\
		                                 &                        & $-$0.17(03)$i$ & $-$0.89(03)$i$ & $-$4.27(13)$i$ & +0.09(01)$i$   & $-$1.83(06)$i$ \\ \hline
		\multirow{2}{*}{$N^{\rm 1}_{1}$} & $0$                   & 12.3(5)      & 105(5)       & 66(2)       & 187(8)       & 43(1)       \\
		                                 & $1$                  & 3.1(2)       & 47(2)       & 21.7(8)       & 215(9)       & 37(1)       \\
		\multirow{2}{*}{$N^{\rm 2}_{1}$} & $0$                   & 5.9(5)       & 49(5)       & 27(3)       & 70(10)       & 20(2)       \\
		                                 & $1$                  & 1.5(2)       & 19(2)       & 9(1)       & 81(12)       & 16(2)       \\
		\multirow{2}{*}{$N^{\rm 3}_{1}$} & $0$                   & 65(2)       & 545(19)       & 316(5)       & 839(17)       & 215(4)       \\
		                                 & $1$                  & 16(1)       & 225(4)       & 106(2)       & 960(20)       & 183(3)       \\
		  \multirow{2}{*}{$N^{1}_{3}$}   & $0$                 & 6.2(3)       & 16(1)       & 56(3)       & 313(16)       & 14(1)       \\
		                                 & $1$                  & 20(6)       & 7.3(4)       & 0.12(1)      & 397(19)       & 1.42(7)       \\
		  \multirow{2}{*}{$N^{2}_{3}$}   & $0$         & 9.2(3)       & 24(1)        & 83(3)        & 467(17)       & 20(2)        \\
		                                 & $1$                  & 31(9)        & 10.9(4)        & 0.17(2)       & 592(21)       & 2.11(8)       \\ \hline\hline
	\end{tabular}
	\caption{The values of parameters for the fit of different partial waves. PW and $I$ represent partial wave and isospin, respectively. The uncertainties of parameters are from MINUIT. $N^i_1~(i=1,2,3)$ denote the normalization factors for datasets from  BES~\cite{BES:2003aic}, CELO~\cite{CLEO:2010fre}, and BESIII~\cite{BESIII:2011aa}, and $N_3^i~(i=1,2)$ correspond to the datasets from BESIII \cite{BESIII:2023vvr,BESIII:2013sbm}. The $N^2_{3}$ unit is $10^2$. The cut-off is $R=1.0$ fm}
	\label{tab:para}
}	
\end{table}
%%%==============================================

In $N\bar{N}$ scattering, the uncertainties of the partial wave amplitudes are estimated mainly from the expected size of the higher order corrections. 
The formulas to obtain theoretical uncertainties of $N\bar{N}$ scattering are~\cite{Epelbaum:2014efa,Dai:2017ont}:
\begin{eqnarray}
\label{Error}
&\Delta\! X\!^{\rm N^3LO}\! (k)\!
=\! \max \! \bigg( \! Q^5\! \Big| X\!^{\rm
    LO}\!(k) \Big|, Q^3 \!\Big|
  X\!^{\rm LO}\!(k) \!-\!  X\!^{\rm NLO}\!(k) \Big|, \nonumber \\
&Q^2 \Big|
  X\!^{\rm NLO}\!(k) \!-\!  X\!^{\rm N^2LO}\!(k) \Big|, Q \Big|
  X\!^{\rm N^2LO}\!(k) \!-\!  X\!^{\rm N^3LO}\!(k) \Big|  \bigg),\nonumber\\
  \label{Eq:error}
\end{eqnarray}
where the expansion parameter $Q$ is defined by
\begin{equation}
\label{expansion}
Q =\max \left( \frac{k}{\Lambda_b}, \; \frac{M_\pi}{\Lambda_b} \right)\,,\nonumber
\end{equation}
with $k$ the momentum and $\Lambda_b$ the breakdown scale.
%%%%
Following Eq.(\ref{Eq:error}) we estimate the uncertainties of the invariant mass spectra coming from the $N\bar{N}$ partial waves, combined together with the ones obtained through bootstrap, where the data is varied within its error by multiplying a normal distribution function \cite{Efron:1979bxm}. They are now exhibited as the orange, dark pink, lime, gold, and cyan bands for $^3P_0$, $^1S_0$, $^3P_1$, $^1D_2$, and $^3P_2$ waves, respectively.

The results for isoscalar waves are in the left column, and the isovector ones are shown in the right column. The invariant mass spectra of the $J/\psi\to\gamma3(\pi^+\pi^-)$ and $J/\psi\to\gamma p\bar{p}$, and the cross section of $p\bar{p}\to 3(\pi^+\pi^-)$, are shown in the first, second, and third row,  in order. 
%%%%
To determine the quantum numbers of the structure around the $\bar{p}p$ threshold (called  $X(1880)$ for simplicity), we follow a simple strategy: the partial wave with the correct quantum numbers should describe all the data well. 

%%%%====================================
%%%%====================================

As can be found from the $3(\pi^+\pi^-)$ invariant mass spectra, the $I=0$ $^3P_0$, $^3P_1$, $^1D_2$ and the $I=1$ $^1D_2$ waves 
give poor description of the data, see the purple dashed, blue dotted, and green dot-dashed lines in the first row of Fig~\ref{fig:3pipi4}. One can exclude these quantum numbers. 
%%%%
From the $p\bar{p}$ invariant mass spectra, one can find that all the waves, except for the $I=0$ $^1S_0$, $^3P_0$ and $I=1$ $^1S_0$ ones, can hardly describe the data. Among them, the $I=0$ $^3P_0$ wave gives a much worse description of the data than the $I=0$ $^1S_0$-wave. 
%%%
From the $p\bar{p}\to 3(\pi^+\pi^-)$ cross section, one can find that only the $I=0$ $^1S_0$ and $I=1$ $^1S_0$ waves can describe the data well. 
This is unsurprising as the amplitudes of $P$- and $D-$ waves are proportional to the momentum $p^L$, resulting in a vanishing cross section at the $\bar{p}p$ threshold. 
%Indeed, even the interference between the higher partial waves and the S-wave will be suppressed near the threshold due to the momentum factor. 
%%%%
In short, only the $I=0$ $^1S_0$ wave survives by describing all the data well; the $ I=1$ $^1S_0$ wave is worse than the isoscalar one but much better than all other waves. Consequently, the $X(1880)$ should be in the $I=0$ $^1S_0$ wave. The $I=1$ $^1S_0$ wave is the second choice, but much worse than the former.  

%%%==============================================
\begin{table}[htbp!]
    \centering
    \begin{tabular}{c|c| c c c c c}
    \hline\hline
     $R$ (fm)   \rule[-0.3cm]{0cm}{8mm}                     & Isospin      & $^1S_0$ & $^3P_0$ & $^3P_1$ & $^1D_2$ & $^3P_2$ \\ \hline
    \multirow{2}{*}{0.9}& $I=0$ & 1.96    & 7.87    &41.51    &67.39    &20.94    \\
                        & $I=1$ & 4.53    &35.09    &30.72    &69.05    &29.97    \\ \hline
    \multirow{2}{*}{1.0}& $I=0$ & 1.99    & 6.74    &33.81    &66.18    &18.58    \\
                        & $I=1$ & 3.30    &31.34    &26.38    &67.82    &25.03    \\ \hline
    \multirow{2}{*}{1.1}& $I=0$ & 2.04    & 6.34    &25.70    &64.78    &16.40    \\
                        & $I=1$ & 3.32    &28.00    &21.20    &66.45    &20.85    \\ \hline 
    \multirow{2}{*}{1.2}& $I=0$ & 2.07    & 6.42    &19.80    &63.24    &14.90    \\
                        & $I=1$ & 3.22    &24.74    &17.13    &64.94    &17.98    \\ \hline\hline                                      
    \end{tabular}
    \caption{The $\chi^2/{\rm d.o.f.}$ for the differential partial waves, where ${\rm d.o.f.}$ represents the number of degrees of freedom.}
    \label{tab:chisqdof}
\end{table} 
% \begin{table}[htbp!]
%     \centering
%     \begin{tabular}{c|c| c c c c c}
%     \hline\hline
%      $R$ (fm)   \rule[-0.3cm]{0cm}{8mm}                     & Isospin      & $^1S_0$ & $^3P_0$ & $^3P_1$ & $^1D_2$ & $^3P_2$ \\ \hline
%     \multirow{2}{*}{0.9}& $I=0$ & 1.09    & 6.43    &28.07    &65.89    &20.69    \\
%                         & $I=1$ & 3.22    &31.16    &26.74    &67.23    &24.56    \\ \hline
%     \multirow{2}{*}{1.0}& $I=0$ & 1.12    & 5.48    &24.27    &64.75    &18.45    \\
%                         & $I=1$ & 2.31    &28.26    &23.59    &66.09    &21.31    \\ \hline
%     \multirow{2}{*}{1.1}& $I=0$ & 1.16    & 5.11    &19.95    &63.44    &16.27    \\
%                         & $I=1$ & 2.34    &25.64    &19.61    &64.81    &18.36    \\ \hline 
%     \multirow{2}{*}{1.2}& $I=0$ & 1.19    & 5.25    &16.41    &61.99    &14.72    \\
%                         & $I=1$ & 2.27    &22.97    &16.13    &63.40    &16.23    \\ \hline\hline                                      
%     \end{tabular}
%     \caption{The $\chi^2/{\rm d.o.f.}$ for the differential partial waves, where ${\rm d.o.f.}$ represents the number of degrees of freedom.}
%     \label{tab:chisqdof}
% \end{table}   
%%%==============================================
To test the stability of our conclusion, we list the $\chi^2/{\rm d.o.f.}$ for different cut-offs, $R=0.9,1.0,1.1,1.2$~fm in Table \ref{tab:chisqdof}. The reasons we choose these cut-offs are as follows: About 1~fm corresponds to the typical scale of the strong interaction; Converting the cut-offs from coordinate space to momentum space gives 438.50, 394.65, 358.78, and 328.88 MeV for 0.9, 1.0, 1.1, and 1.2~fm, respectively. These cut-offs are widely applied in studying the $NN,N\bar{N}$ interaction within ChEFT \cite{Epelbaum:2014efa,Dai:2017ont}. 
As can be found, the $I=0$ $^1S_0$ wave has the smallest $\chi^2/{\rm d.o.f.}$.  
The $I=1$ $^1S_0$ wave has the second smallest $\chi^2/{\rm d.o.f.}$, from 3.2 to 4.5. All other waves have much larger $\chi^2/{\rm d.o.f.}$. This confirms that the $X(1880)$ most likely has the quantum numbers $I~J^{PC}=0~0^{-+}$.
Indeed, we have also tested the present analysis within $T_{\rm Lab}\lesssim 100$~MeV, and the same conclusion is obtained. 

Further, we calculated the p-values of the linear correlation between the fit results and the data sets.
As is known, the p-value should be above 0.05 to have confidence that the test passes. In our analysis, the theoretical prediction also has an uncertainty, which should be included. For each data point, the variance is defined as $\sigma_{{\rm tot}}=\sqrt{\sigma^2_{\rm exp}+\sigma^2_{\rm theo}}$~\cite{Bevington:2002}, where $\sigma_{\rm exp}$ is the error from experiment and $\sigma^2_{\rm theo}$ is from theory.
Our p-values are listed in Table~\ref{tab:pvalue}. As can be found, only the $I=0$ $^1S_0$ partial wave passes the test, after removing three data points that are far away from other data points, i.e., the red points at $E_{\rm cm}=$1.89, 1.93~GeV of the invariant mass spectra for $J/\psi\to\gamma 3(\pi^+\pi^-)$, shown in the first row graphs of Fig.~\ref{fig:3pipi4}, and the blue point at $M_{p\bar{p}}-2m_p$=0.018~GeV of the invariant mass spectra for $J/\psi\to\gamma p\bar{p}$, as shown in the second row graphs of Fig.~\ref{fig:3pipi4}. After removing these data points, one finds $\chi^2/{\rm d.o.f}=1.12$ with the cut-off 1~fm for the $I=0$ $^1S_0$ wave, and its p-value is $0.215$. Hence, the quantum numbers of $X(1880)$ are most likely  $IJ^{PC}=00^{-+}$. The p-values of the isovector $^1S_0$ wave is less than $10^{-7}$, and other partial waves are smaller than $10^{-34}$.
 %%%%%%%%%%%%%%%
%%%==============================================
\begin{table}
	\centering
	\begin{tabular}{c|c| c c c c c }
		\hline\hline
		$R$ (fm)   \rule[-0.3cm]{0cm}{8mm} & Isospin & $^1S_0$ & $^3P_0$ & $^3P_1$ & $^1D_2$ & $^3P_2$ \\ \hline
		       \multirow{2}{*}{0.9}        &  $I=0$  & 0.261   & 0.000   & 0.000   & 0.000   & 0.000   \\ \cline{2-7}
		                                   &  $I=1$  & 0.000   & 0.000   & 0.000   & 0.000   & 0.000   \\ \hline
		       \multirow{2}{*}{1.0}        &  $I=0$  & 0.215   & 0.000   & 0.000   & 0.000   & 0.000   \\ \cline{2-7}
		                                   &  $I=1$  & 0.000   & 0.000   & 0.000   & 0.000   & 0.000   \\ \hline
		       \multirow{2}{*}{1.1}        &  $I=0$  & 0.168   & 0.000   & 0.000   & 0.000   & 0.000   \\ \cline{2-7}
		                                   &  $I=1$  & 0.000   & 0.000   & 0.000   & 0.000   & 0.000   \\ \hline
		       \multirow{2}{*}{1.2}        &  $I=0$  & 0.137   & 0.000   & 0.000   & 0.000   & 0.000   \\ \cline{2-7}
		                                   &  $I=1$  & 0.000   & 0.000   & 0.000   & 0.000   & 0.000   \\ \hline\hline
	\end{tabular}
	\caption{The p-value for the differential partial waves and isospin for cut-offs in the range 0.9-1.2 fm. }
	\label{tab:pvalue}
\end{table}
%%%==============================================
%%%%%%%%%%%%%%%%%%%%%%%%%%%%%%%

\begin{figure}[htpb!]
 \centering
 \includegraphics[width=0.9\linewidth]{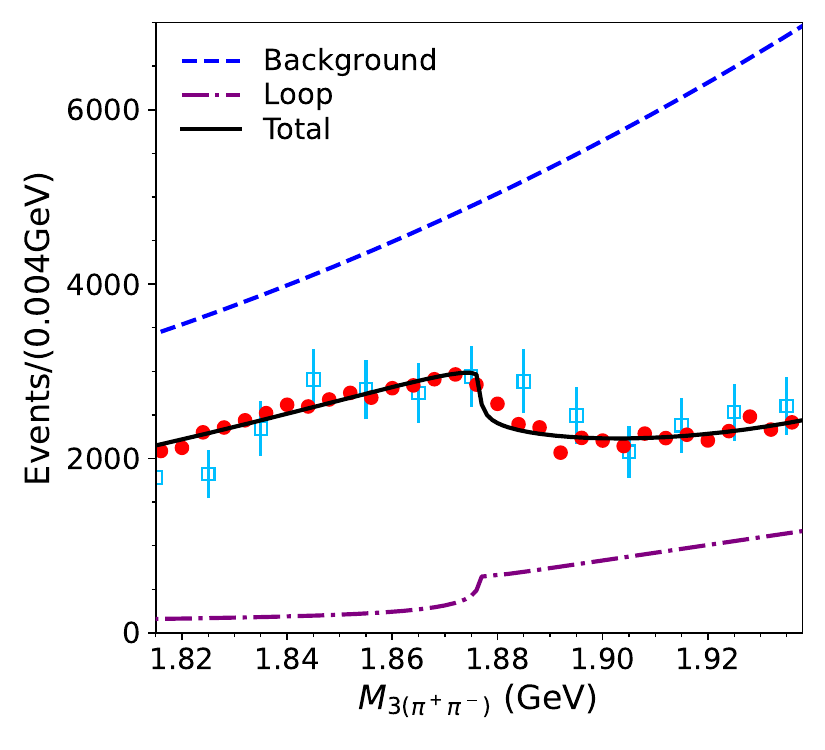}
 \caption{Contributions from different parts to the $J/\psi\to\gamma3(\pi^+\pi^-)$ spectra. The solution is an isoscalar $^1S_0$ wave. The black solid, blue dashed, and purple dash-dotted lines indicate the contribution of the total, Born term (background), and the loop effect from $N\bar{N}$ rescattering, respectively.}
 \label{fig:3pipidevide}
\end{figure}
%%%==============================================
%In statistics, the threshold values for p-values are typically set at 0.05 and 0.01. If p$<$0.01, it indicates a strong correlation, while 0.01$<$p$<$0.05 indicates a weak correlation. 

After fixing the quantum numbers, one can analyze the contributions to the structure around the threshold from different parts as shown in Fig.~\ref{fig:3pipidevide}. 
The Born term/background and loop contributions, corresponding to the first and second terms of Eq.~(\ref{Eq:AmpF}), are shown as blue dashed and purple dash-dotted lines, respectively. As can be found, the structure in the $3(\pi^+\pi^-)$ variant mass spectrum is from the interference between the loop and background contributions. The background is smooth, and only the loop contribution, i.e., the $N\bar{N}$ rescattering effects, generate a structure that is of a cusp-like shape rather than a Breit-Wigner shape. 
It should be pointed out that from the $N\bar{N}$ scattering partial wave amplitudes, one can not find any poles around the threshold compatible with the cusp-like  shape of the loop contribution. This excludes the glueball case and supports that the $X(1880)$ is generated by the threshold behavior.

\section{Summary}
    In this work, we have presented  a combined analysis on the decay process of  the $J/\psi\to\gamma 3(\pi^+\pi^-)$ and $J/\psi\to\gamma p\bar{p}$ reactions and the scattering process of $p\bar{p}\to 3(\pi^+\pi^-)$. The distorted wave Born approximation is applied to solve the amplitudes, with input of the $N\bar{N}$ scattering amplitudes given by chiral effective field theory up to N$^3$LO. 
The isoscalar and isovector $^1S_0$, $^3P_0$, $^3P_1$, $^1D_2$ and $^3P_2$ partial waves are taken into the analysis one by one. 
%%%%%
Our result suggests that the quantum numbers of the structure around the $\bar{p}p$ threshold, the $X(1880)$, discovered in the $3(\pi^+\pi^-)$ invariant mass spectrum of $J/\psi\to\gamma 3(\pi^+\pi^-)$ process, is most likely to be $I\,J^{PC}=0\,0^{-+}$. 
Further, the $X(1880)$ is generated by the $\bar{N}N$ threshold effect. 
This can be checked by the future experiments.

\section*{Acknowledgements}
	\label{Sec:V}
This work is supported by the National Natural Science Foundation of China (NSFC) with Grants No.~12322502, 12447186, 12335002, Joint Large Scale Scientific Facility Funds of the NSFC and Chinese Academy of Sciences (CAS) under Contract No.~U1932110, Hunan Provincial Natural Science Foundation with Grant No. 2024JJ3004, Fundamental Research Funds for the central universities, 
and by the MKW NRW under the funding code NW21-024-A and by ERC EXOTIC (grant No.~101018170).
The work of UGM was supported in part by the CAS President's International Fellowship Initiative (PIFI) (Grant No.~2025PD0022).

\bibliography{ref.bib}

\end{document}